\RequirePackage{fix-cm}
\documentclass[smallcondensed]{svjour3ujm}
\smartqed
\usepackage{graphicx}
\usepackage[authoryear]{natbib}
\newcommand{\braket}[2]{\langle#1|#2\rangle}

\newcommand{\ket}[1]{|#1\rangle}
\newcommand{\bq}{\begin{quote}}
\newcommand{\eq}{\end{quote}}
\newcommand{\ben}{\begin{enumerate}}
\newcommand{\een}{\end{enumerate}}
\newcommand{\be}{\begin{equation}}
\newcommand{\ee}{\end{equation}}
\begin{document}
\title{Quantum Mechanics in a New Light}
\author{Ulrich J. Mohrhoff}
\institute{Ulrich J. Mohrhoff\\
Sri Aurobindo International Centre of Education\\Pondicherry 605002 India\\
Tel.: +91-93454-21357\\
\email{ujm@auromail.net}\\
\\
\textbf{The final publication is available at\\
http://link.springer.com/article/10.1007/s10699-016-9487-6\\
\\
Published in Foundations of Science DOI 10.1007/s10699-016-9487-6\\
}}
\maketitle
\begin{abstract}
Although the present paper looks upon the formal apparatus of quantum mechanics as a calculus of correlations, it goes beyond a purely operationalist interpretation. Having established the consistency of the correlations with the existence of their correlata (measurement outcomes), and having justified the distinction between a domain in which outcome-indicating events occur and a domain whose properties only exist if their existence is indicated by such events, it explains the difference between the two domains as essentially the difference between the manifested world and its manifestation. A single, intrinsically undifferentiated Being manifests the macroworld by entering into reflexive spatial relations. This atemporal process implies a new kind of causality and sheds new light on the mysterious nonlocality of quantum mechanics. Unlike other realist interpretations, which proceed from an evolving-states formulation, the present interpretation proceeds from Feynman's formulation of the theory, and it introduces a new interpretive principle, replacing the collapse postulate and the eigenvalue--eigenstate link of evolving-states formulations. Applied to alternatives involving distinctions between regions of space, this principle implies that the spatiotemporal differentiation of the physical world is incomplete. Applied to alternatives involving distinctions between things, it warrants the claim that, intrinsically, all fundamental particles are identical in the strong sense of numerical identical. They are the aforementioned intrinsically undifferentiated Being, which manifests the macroworld by entering into reflexive spatial relations.
\end{abstract}

\section{Introduction}\label{sec:intro}
While the stunningly elegant formalism of quantum mechanics can be written down on a napkin, attempts to make sense of it fill entire libraries \citep[p.~19]{Schlosshauer2011}. Not a single reliable experiment or observation has ever been at odds with what it predicts. Its ever-growing range of technological applications borders on magic, yet no one seems to know how the magic works. Subatomic particles, atoms, and molecules have ways of behaving available to them which seem to be quite unlike what we know how to think about.

The difficulty is compounded by several factors. One of these factors is the way in which quantum mechanics is routinely taught. While a junior-level classical mechanics course devotes a considerable amount of time to different formulations of classical mechanics---such as Newtonian, Lagrangian, Hamiltonian, least action---even graduate-level courses emphasize one particular formulation of quantum mechanics almost to the exclusion of all variants. There are at least nine formulations of quantum mechanics \citep{Styeretal}, among them Heisenberg's matrix formulation, Schr\"odinger's wave-function formulation, Feynman's path-integral formulation, Wigner's phase-space formulation, and the density-matrix formulation. It would seem reasonable to think that the interpretation of quantum mechanics should be based on features that are common to all formulations of the theory, not on the mathematical idiosyncracies of a particular formulation. This point may be illustrated by an analogy with the inertial frames of the special theory of relativity. If the same physical situation is described using different frames, certain features (like the temporal order of events connected by light signals) are common to all descriptions, while other features (like the temporal order of events that are simultaneous in \emph{some} frames) are not. Just as it is standard practice to consider only those features objective that are shared by all relativistic descriptions of a physical situation, so it should be standard practice to extract the physical import of the mathematical apparatus of quantum mechanics from those features that are common to all formulations of the theory. What is common to all formulations is that they provide us with tools for calculating statistical correlations between measurement outcomes, or with algorithms for assigning probabilities to possible measurement outcomes on the basis of actual measurement outcomes.%
\footnote{\label{oo}There is further evidence that quantum mechanics is essentially a probability calculus or a calculus of correlations. If one accepts the existence of ``ordinary'' objects, defined as objects that (i)~have spatial extent (they ``occupy'' space), (ii)~are composed of finite numbers of objects that lack spatial extent, and (iii) neither collapse nor explode as soon as they are formed, then one needs the uncertainty principle to stabilize these objects, and the ``uncertainties'' $\Delta x$ and $\Delta p$ must be measures of an objective indeterminacy or fuzziness. This is because what ``fluffs out'' ordinary objects cannot be anyone's ignorance of the exact values of their internal relative positions and momenta; it can only be an objective indeterminacy. We describe this objective indeterminacy by assigning probabilities to the possible outcomes of measurements. If we then look for an appropriate probability calculus---for how the probabilities of the possible outcomes of a measurement depend on actual outcomes, and how they depend on the time of the measurement to the possible outcomes of which they are assigned---what we find is the formal apparatus of quantum mechanics. This argument is set forth in \citep{Mohrhoff2009b} and in greater detail in \citep[Chapter 8]{Mohrhoff2011}.}

In \emph{The Ashgate Companion to Contemporary Philosophy of Physics} \citep{Wallace}, a distinction is drawn between the Bare Quantum Formalism, which is said to precede any notion of probability and measurement, and the Quantum Algorithm, which serves to extract the probabilities of measurement outcomes: ``The Bare Quantum Formalism \dots\ is an elegant piece of mathematics; the Quantum Algorithm is an ill-defined and unattractive mess.''  This ``ill-defined and unattractive mess'' has come to be seen as ``the great scandal of physics.'' Otherwise it is known as the Measurement Problem. What is perceived as particularly offensive is the apparently inevitable invocation of primitives like ``measurement'' and ``observer,'' which are beyond the scope of the Bare Quantum Formalism. But if it is possible to derive the theory by an argument that is transcendental in the sense of Kant, as a precondition of the possibility of ordinary objects (defined in note \ref{oo}), and if this condition of possibility turns out to be a probability calculus, then the distinction between a Bare Quantum Formalism (which precedes the notions of probability and measurement) and a Quantum Algorithm (which serves to extract probabilities from the former) is misconceived. There is no formalism that precedes the notions of probability and measurement. Every piece of the formalism can be derived, and therefore can be understood, as an aspect of a probability calculus.%
\footnote{The notion that a Bare Quantum Formalism can be separated from the quantum-mechanical probability calculus is also suggested by the manner in which the axioms of standard quantum mechanics are routinely stated. This is because the first several axioms tend to be stated without explicit reference to probabilities; only the last couple of axioms refer to probabilities. Yet every single axiom makes perfect sense only as a feature of a probability calculus, as has been argued in \citep{Mohrhoff2009b} and \citep[Section 16.1]{Mohrhoff2011}.}

The probabilities that quantum mechanics allows us to calculate have two obvious and natural dependencies. They depend on the actual outcomes on the basis of which they are assigned, and they depend on the times of the measurements to the possible outcomes of which they are assigned. But since it is widely believed (owing to how the theory is routinely taught) that the aim of interpreting quantum mechanics is to make sense of Schr\"odinger's wave-function formulation or Dirac's cognate state-vector formulation, the two dependencies are now almost universally understood as alternating ways in which quantum states evolve: between measurements, a quantum state changes continuously and predictably, whereas at the time of a measurement it changes (or appears to change) discontinuously and unpredictably. This way of thinking engendered a more specific problem---the problem of objectification, whose first rigorous formulations are due to \citet{vonNeumann} and \citet{Pauli1933}. Detailed recent discussions are \citep{Buschetal} and \citep{Mittelstaedt}. 

Every discussion of this problem begins by describing the process of measurement as comprising three stages: the system preparation, a continuous dynamical process called ``premeasurement''  ($pm$), and the seemingly miraculous appearance of an outcome called ``objectification'' ($ob$). In terms of projection-valued measures:
\begin{equation}
\sum_{k}c_k\ket{A_0}\ket{q_k}\stackrel{(pm)}{\longrightarrow}
\sum_{k}c_k\ket{A_k}\ket{q_k}\stackrel{(ob)}{\longrightarrow}
\ket{A(q)}\ket{q}.
\label{eq_pmob}
\end{equation}
Every interpretation of quantum mechanics that attaches to a quantum state an ontological significance which goes beyond that of a probability algorithm, founders over the transition from the state that is supposed to complete the premeasurement to the state that is supposed to complete the measurement, in which $q$ (one of the possible outcomes $q_k$) is indicated. \citet[Sect.~4.3b]{Mittelstaedt} and \citet[Sect.~III.6.2]{Buschetal} have even proved theorems to the effect that the objectification problem is insoluble. The least that can be said is that such realist interpretations raise issues that are far from being settled. From an operationalist or antirealist point of view, this does not come as a surprise.  For the wave function's dependence on time is not the continuous time-dependence of an evolving physical state. The time it depends on is the time of the measurement to the possible outcomes of which it serves to assign probabilities. As \citet{Peres84} has stressed, ``there is no interpolating wave function giving the `state of the system' between measurements''.%
\footnote{Here Peres echoes Bohr's insistence that what happens between the preparation of a system and a measurement is a holistic phenomenon, which cannot be decomposed into the unitary evolution of a quantum state and a subsequent ``collapse'' of the same: ``all unambiguous interpretation of the quantum mechanical formalism involves the fixation of the external conditions, defining the initial state of the atomic system concerned and the character of the possible predictions as regards subsequent observable properties of that system. Any measurement in quantum theory can in fact only refer either to a fixation of the initial state or to the test of such predictions, and it is first the combination of measurements of both kinds which constitutes a well-defined phenomenon.'' \citep{Bohr1939}}
Nothing good can come out of the transmogrification of a probability algorithm into an evolving physical state. 

During a successful measurement, the apparatus does indeed evolve from a neutral state, in which it is ready to perform its function, to a state in which it indicates an outcome, but this evolution does not involve the unitary ``evolution'' of the quantum state of a composite quantum system into a superposition of the form $\sum_k c_k\ket{a_k}\ket{b_k}$, where $\ket{a_k}$ and $\ket{b_k}$ are eigenstates of observables $A$ and~$B$. This superposition ``obtains'' at a time $t$ only in the conditional sense that a joint measurement of $A$ and $B$, made at the time~$t$, would (or will) indicate the pair of outcomes $a_k$ and $b_k$ with probability $|c_k|^2$, and that it would (or will) indicate the pair of outcomes $a_i$ and $b_k$ with probability~0 if $i\neq k$. The strict correlation between the possible outcomes of the two observables warrants another conditional statement, to wit: if (say) a measurement of $A$ yields $a_k$, then a measurement of $B$ would (or will) yield $b_k$ with probability~1. What it does not warrant is the claim that $B$ then \emph{has} the value~$b_k$. Probability 1 is not sufficient for ``is'' or ``has''.%
\footnote{Another misconception muddying the interpretive waters is the so-called eigenvalue-eigenstate link, a postulate that \citet[pp.~46--47]{Dirac} formulated thus: ``The expression that an observable `has a particular value' for a particular state is permissible \dots\ in the special case when a measurement of the observable is certain to lead to the particular value, so that the state is an eigenstate of the observable.'' If the time a quantum state depends on is the time of the measurement to the possible outcomes of which it serves to assign probabilities, then the time at which an observable has a particular value can only be the time of an actual measurement.}

What the superposition $\sum_{k}c_k\ket{A_k}\ket{q_k}$ actually tells us, therefore, is this: if the apparatus were subjected to a measurement, and if the outcome were $A_k$, a measurement of the system observable $Q$ would yield $q_k$ with probability~1. What is \emph{logically inconsistent}, what is \emph{self-contradictory}, is to read $A_k$ here as representing an outcome-indicating property. As a probability calculus, quantum mechanics presupposes the events to which it serves to assign probabilities. It presupposes the existence of outcome-indicating devices with outcome-indicating properties. But no device can (i) be an outcome-indicating device and (ii) enter a superposition involving outcome-indicating properties. No measurement apparatus can (i)~serve as a measurement apparatus and (ii)~enter a state that disqualifies it as a measurement apparatus---unless one embraces a many-worlds interpretation, in which every outcome-indicating state is realized in a separate world.%
\footnote{Many-worlds interpretations, like other realist interpretations, face a number of issues that (to say the least) have by no means been conclusively resolved \citep{Barrett,Marchil2015}.}

Because quantum mechanics presupposes the events to which it serves to assign probabilities, it cannot be called upon to account for their existence. But of course it has to be consistent with it. Any argument or way of thinking which leads to the conclusion that the quantum-mechanical correlation laws are inconsistent with the existence of their correlata will of necessity depend on at least one false assumption. Insofar as there is a measurement problem, it can be described as the task of demonstrating the theory's semantic consistency---or else the task of identifying the false assumption(s) underlying the theory's perceived lack of semantic consistency. The term ``semantic consistency'' was introduced by von Weizs\"acker. By the semantic consistency of a theory he meant ``that its preconceptions, how we interpret the mathematical structure physically, will themselves obey the laws of the theory'' \citep[p.~260]{vW}. Among the obvious preconceptions of quantum mechanics are the existence of outcome-indicating devices having outcome-indicating properties and being able to evolve from a neutral state to an outcome-indicating state. Section \ref{sec:semantic} takes up the task of demonstrating that these preconceptions of the theory do indeed obey the laws of the theory. In particular, it will be shown how macroscopic objects and outcome-indicating properties can and should be defined so that the evolution of measurement devices from neutral states to outcome-indicating states is consistent with the unitary dependence of probabilities on the time of the measurement to the possible outcomes of which they are assigned.

The demonstration of quantum theory's semantic consistency proceeds in two steps. First it is shown  that the spatial differentiation of the physical world does not go ``all the way down'': if conceptually we keep partitioning space into smaller and smaller regions, we reach a point beyond which the distinctions we make between regions no longer correspond to anything in the actual physical world. Subsequently it is shown that the incompleteness of the spatial differentiation of the physical world implies the existence of a non-empty class of objects whose positions are ``smeared out'' only relative to an imaginary spatiotemporal background that is more differentiated than the physical world. This paves the way for rigorous definitions of ``macroscopic object'' and ``outcome-indicating event'' in the theory's own terms.

Despite its insistence on the characterization of quantum mechanics as a probability calculus, the present paper aims to provide a coherent picture of reality. A realist interpretation of the theory does not have to be a realist interpretation of a universal state vector or wave function. The reason why a coherent picture of reality has not yet been found may well be that the realist interpretations on offer treat quantum states as evolving physical states. At any rate, if the aim is to arrive at a coherent picture of reality, evolving-states formulations of the theory cannot be relied on. Section \ref{sec:newintpl} makes use of Feynman's formulation \citep{FHS} to introduce an interpretive principle that takes the place of the collapse (or projection) postulate of the former formulations. Instead of attempting to account for the objectification of the values of measured observables, this interpretive principle specifies the conditions under which distinctions that we tend to make can \emph{not} be objectified (i.e., can \emph{not} be represented as real differences in the actual physical world).

In Section \ref{sec:idpas} this interpretive principle is applied to two paradigmatic setups, one concerning distinctions between regions of space, the other concerning distinctions between things. The first application confirms the conclusion, reached in Section \ref{sec:semantic}, that space cannot be an intrinsically differentiated expanse. The so-called parts of space need to be physically realized by detectors. The second application is exemplified by a scattering experiment. Two identical particles---particles lacking individualizing properties---are initially found to be moving northward and southward, respectively. The next thing we know is that the same two particles are found to be moving eastward and westward, respectively. The question then is: which incoming particle is identical with which outgoing particle? It  is well known that there is no answer to this question. The distinction we make between the two possible identifications cannot be objectified. What does this tell us about the physical world? The proposed answer is simple, straightforward, but also highly counterintuitive. It is that the incoming particles (and therefore the outgoing ones as well) are \emph{one and the same entity}. What's more, there is no compelling reason to believe that this identity ceases when it ceases to have observable consequences owing to the presence of individuating properties.  We are free to take the view that \emph{intrinsically} each particle is numerically identical with every other particle. What presents itself here and now with these properties and what presents itself there and then with those properties is one and the same entity, to which I shall refer as ``Being.''

Another reason we find it so hard to make sense of the quantum theory is that it answers a question we are not in the habit of asking. Instead of asking what the ultimate constituents of matter are and how they interact and combine, we should  ask: how are forms manifested? To this question, too, a simple, straightforward, and counterintuitive answer can be given: the shapes of things resolve themselves into reflexive spatial relations---that is, \emph{self}-relations, relations between Being and the self-same Being. By entertaining reflexive spatial relations, Being supports (i)~what looks like a multiplicity of relata if the reflexive quality of the relations is ignored, and (ii)~what looks like a substantial expanse if the spatial quality of the relations is reified. These ideas, put forth in Section~\ref{Manifestation}, pave the way for a deeper understanding of the nature of the quantum domain and its relation to the ``classical'' or macroscopic domain, beyond the linguistic necessity of speaking about the quantum domain in terms of correlations between macroscopic events, which was stressed by \citet{Bohr1934}. The distinction between the two domains now presents itself as essentially a distinction between the \emph{manifested world} and its \emph{manifestation}. 

The penultimate section addresses the mysterious nonlocality of quantum mechanics. The manifestation of the spatiotemporal world---the atemporal process by which Being enters into reflexive relations and matter and space come into being as a result---is the nonlocal event \emph{par excellence}. It is the root of all quantum correlations, not only the seemingly inexplicable ones between simultaneous events in different locations but also the seemingly explicable ones between successive events in the same location. This section concludes with a provocative thought. What if the force at work in the physical world---the dynamism by which Being manifests the physical world---were an infinite (unlimited) force operating under self-imposed constraints? In that case we would have no reason to be surprised or dismayed by the impossibility of explaining the quantum-mechanical correlation laws in terms of physical mechanisms, inasmuch as it would be self-contradictory to invoke a physical mechanism to explain the working of an infinite force. What would need explaining is why this force works under the particular constraints that it does. That, of course, is a question for pure metaphysics.%
\footnote{If the existence of ``ordinary'' objects (as defined in note \ref{oo}) is a fact, then a teleological or anthropic demonstration of what is entailed by it can go a long way towards explaining why the laws of physics have the form that they do \citep{Mohrhoff2002,Mohrhoff2009b, Mohrhoff2011}. The question for pure metaphysics is then why ``ordinary'' objects, which have spatial extent, are composed of finite numbers of objects lacking spatial extent. For a possible answer see the final section and \citep{Mohrhoff2014a}.}
The final section, prompted by comments by a reviewer, briefly addresses the relation between quantum mechanics and consciousness.

\section{Semantic consistency---the macroworld}\label{sec:semantic}
While quantum mechanics can tell us that the probability of finding a particle in a given region of space is~1, it is incapable of  giving us a region of space. For this a detector is needed. A detector is needed not only to indicate the presence of a particle in a region but also---and in the first place---to physically realize a region, so as to make it possible to attribute to a particle the property of being inside. Speaking more generally, a macroscopic apparatus is needed not only to indicate the possession of a property by a quantum system but also---and in the first place---to make a set of properties available for attribution to the system. In addition, a macroscopic clock is needed to realize attributable times. This, of course, is vintage Bohr, who insisted that the ``procedure of measurement has an essential influence on the conditions on which the very definition of the physical quantities in question rests'' \citep{Bohr1935}. 

But if detectors are needed to realize regions of space, space cannot be intrinsically partitioned. If we conceive of it as partitioned, we can do so only as far as regions of space can be realized---i.e., to the extent that the requisite detectors are physically possible. This extent is limited by the indeterminacy principle, inasmuch as this rules out the existence of detectors with arbitrarily small sensitive regions that are (and remain) sharply localized relative to each other. If such regions cannot be realized (as the sensitive regions of detectors), then they are not available for attribution (as positions). Hence, if conceptually we keep partitioning space into smaller and smaller regions, we will reach a point beyond which the distinctions we make between regions no longer correspond to anything in the actual physical world. We can conceive of a partition of the physical world into \emph{finite} regions so small that none of them can be attributed (as a position) because none of them is available for attribution. In other words, physical space cannot be realistically modeled as an actually existing manifold of intrinsically distinct points. In yet other words, the spatial differentiation of the physical world is incomplete---it does not go ``all the way down.'' 

The same goes for the world's temporal differentiation, and this not only because of the relativistic interdependence of distances and durations. Just as the properties of quantum systems or the values of quantum observables need to be realized---made available for attribution---by macroscopic devices, so the times at which properties or values are possessed need to be realized by macroscopic clocks. And just as it is impossible for macroscopic devices to realize sharp positions, so it is impossible for macroscopic clocks to realize sharp times~\citep{Hilgevoord98}. Hence, neither the spatial nor the temporal differentiation of the physical world goes ``all the way down.''

In a world that is incompletely differentiated spacewise, the next best thing to a sharp trajectory is a trajectory that is so sharp that the bundle of sharp trajectories over which it is statistically distributed is never probed. In other words, the next best thing to an object with a sharp position is an object whose position probability distribution is and remains so narrow that there are no detectors with narrower position probability distributions---detectors that could probe the region over which the object's position extends. If the spatiotemporal differentiation of the physical world does not go ``all the way down,'' such objects must exist. If I call them ``macroscopic objects,'' and if I call their positions ``macroscopic positions,'' it is not intended to mean that they are so large and/or massive as to behave like classical objects FAPP (for all practical purposes), but in the more rigorous sense just spelled out. By the ``macroworld'' I shall mean the totality of macroscopic positions.

What can be deduced from this characterization of macroscopic positions is that the events by which their values are indicated are (diachronically) correlated in ways that are consistent with the laws of motion that quantum mechanics yields in the classical limit. For any given time $t$ the following holds: if every event that indicates a macroscopic position prior to the time $t$ were taken into account, then---given the necessarily finite accuracy of position-indicating events---every event that indicates a macroscopic position at a later time would be consistent with all earlier position-indicating events and the classical laws of motion. There is, however, one necessary exception: in order to permit a macroscopic object---the proverbial pointer---to indicate the value of an observable, its position must be allowed to change unpredictably if and when it serves to indicate an outcome.

Macroscopic objects thus follow trajectories that are only counterfactually indefinite. Their positions are ``smeared out'' only in relation to an imaginary spatiotemporal background that is more differentiated than the physical world. No value-indicating event reveals the indefiniteness of a macroscopic position (in the only way it could, through a departure from what the classical laws predict). The testable correlations between outcomes of measurements of macroscopic positions are therefore consistent with both the classical and the quantum laws.%
\footnote{If the indicated values of a supposedly macroscopic position turned out to be inconsistent with a classical law of motion, then this particular position would not actually be a macroscopic position.}
That makes it possible to attribute to macroscopic positions a measurement-independent reality, and this makes it possible for macroscopic positions to define the obtainable values of observables and to indicate the outcomes of measurements.%
\footnote{For extended versions of the present line of reasoning see \citep{Mohrhoff2009a,Mohrhoff2014}.}

According to a theorem due to \citet{CH}, relativistic quantum theory engenders a fundamental conflict between causality and localizability, so that particle talk is ``strictly fictional'': 
\bq
The argument for localizable particles appears to be very simple: Our experience shows us that objects (particles) occupy finite regions of space. But the reply to this argument is just as simple: These experiences are illusory! Although no object is strictly localized in a bounded region of space, an object can be well-enough localized to give the appearance to us (finite observers) that it is strictly localized.
\eq
What Clifton and Halvorson have established is in some respects analogous to the non-objectification theorems proved by \citet[Sect.~4.3(b)]{Mittelstaedt} and the insolubility theorem for the objectification problem due to \citet[Sect.~III.6.2]{Buschetal}. While (according to these theorems) non-relativistic quantum mechanics cannot account for the measurement outcomes which it presupposes and serves to correlate, relativistic quantum mechanics cannot account for the particle detections which it presupposes and serves to correlate. The latter provides us with conditional statements of the following form: if a set of particles ${\cal P}_i$ with momenta $p_i$ come together in a scattering event, then such is the probability with which a set of particles ${\cal Q}_k$ with momenta $q_k$ will emerge from the event. The theory requires that in-states be prepared and out-states be observed, but it leaves the operational implementation to the experimenters. Experimenters use a generalized version of Bohr's correspondence principle to identify these states with the particle types and momenta they obtain by analyzing particle tracks, and it is these data that the theory serves to correlate.%
\footnote{According to \citet[p. XII]{Falkenburg}, ``quantum mechanics and quantum field theory only refer to individual systems due to the ways in which the quantum models of matter and subatomic interactions are linked by semi-classical models to the classical models of subatomic structure and scattering processes. All these links are based on tacit use of a generalized correspondence principle in Bohr's sense (plus other unifying principles of physics).'' This generalized correspondence principle, due to \citet{Heisenberg}, serves as ``a semantic principle of continuity which guarantees that the predicates for physical properties such as `position', `momentum', `mass', `energy', etc., can also be defined in the domain of quantum mechanics, and that one may interpret them operationally in accordance with classical measurement methods. It provides a great many inter-theoretical relations, by means of which the formal concepts and models of quantum mechanics can be filled with physical meaning'' \citep[p.~191]{Falkenburg}.}

Whether or not particles are localizable depends on our answer to the question: localizable relative to what? How are the positions of particles defined---the \emph{observable} positions, the positions that we can meaningfully attribute to particles? The most basic axiom of field theory---so basic that it is rarely explicitly stated---postulates the existence of a spatiotemporal manifold~$\cal M$. What Clifton and Halvorson have shown is that a particle cannot be in a state in which the probability of finding it within any finite spatial region of $\cal M$ equals~1. $\cal M$, however, is not where experiments are performed. The possible outcomes of position measurements are not defined relative to a completely differentiated spatial background such as~$\cal M$. Attributable positions are defined by the sensitive regions of real-world detectors, which also cannot be localized in any finite spatial region of $\cal M$. What is strictly fictional, therefore, is the spacetime manifold postulated by quantum field theory. What Clifton and Halvorson have shown is not that there are no localizable particles but that this manifold is not localizable relative to the positions that can be meaningfully attributed to particles. 

As long as one takes detection events to be localized relative to~$\cal M$,  arguments based on the quantum-mechanical probability calculus can establish the theory's semantic consistency only FAPP (for all practical purposes). This applies in particular to the most vigorously pursued strategy for demonstrating the consistency of the quantum correlations with the existence of their correlata, which capitalizes on the phenomenon of environment-induced decoherence \citep{Joosetal,Zurek,Schlosshauer2007}.  If detection events are localized relative to real-world detectors, on the other hand, decoherence arguments contribute to showing that macroscopic objects (as previously defined) exist (and this not just FAPP), by making the existence of such objects overwhelmingly likely.

If quantum states---probability algorithms such as state vectors and wave functions---are mistaken for evolving physical states, and if their dependence on the time of a measurement is mistaken for a natural process that governs the evolution of these physical states, then collapse and objectification, which seem to be required to conclude the measurement process, become preternatural indeed. It therefore does not come as a surprise that \citet{vonNeumann} himself, the first to codify the tripartite measurement scheme (system preparation---unitary premeasurement---objectification), was led to conclude that the final termination of a measurement is in the conscious observer. The interpretation presented here has no need to invoke conscious observers, not even in the manner of many-worlds interpretations, which multiply the number of conscious observers in the Universe by the number of possible outcomes each time a measurement is made. Because macroscopic positions possess a measurement-independent reality, which allows them to define the values of observables and to indicate the outcomes of measurements, quantum mechanics does not require reference to conscious observers any more than classical mechanics does. What quantum mechanics cannot dispense with is reference to measurements (outcome-indicating events).

\section{A new interpretive principle}\label{sec:newintpl}
It is customary to distinguish between interpretations of quantum mechanics that aim to present a coherent picture of reality, and anti-realist (or empiricist, or operationalist) interpretations, which look upon the theory's formal apparatus as a tool for making predictions. It is rarely (if ever) appreciated that these alternatives are not exclusive. There are other ways of presenting a coherent picture of reality than to invest with physical reality a mathematical feature of a particular formulation of quantum mechanics such as the wave function. There may be a coherent picture of reality that explains why the formal apparatus of quantum mechanics is essentially a probability calculus, and why quantum states cannot be regarded as evolving physical states. It may well be because the realist interpretations on offer---the leading contenders being many-worlds \citep{Barrett,Saundersetal}, the de Broglie-Bohm theory \citep{Bohm1952}, and dynamical collapse theories \citep{GRW,Pearle1989}---treat quantum states as evolving physical states, that a coherent picture of reality has not yet been found.

The present paper aims to present such a picture. To this end, the most useful formulation of quantum mechanics is Feynman's \citep{FHS}, not least because all it essentially does is to assign probability amplitudes to alternatives, which are defined as sequences of measurement outcomes, or as the continuum limits of such sequences. 

Both the wave-function formulation and Feynman's feature a pair of dynamical principles. In the former they are unitary evolution and collapse/objectification, in the latter they are summation over amplitudes (followed by taking the absolute square of the sum) and summation over probabilities (preceded by taking the absolute square of each amplitude). In the context of the wave-function formulation, unitary evolution seems natural; what calls for explanation is collapse (the projection postulate) and objectification (postulated via the eigenvalue--eigenstate link). In the context of Feynman's formulation, adding probabilities seems natural since this conforms to classical probability theory; what calls for explanation is why we have to add amplitudes. What is at issue, therefore, is not what causes wave functions to collapse and outcomes to exist but why we have to add amplitudes whenever quantum mechanics requires us to do so. To answer this question, the following interpretive principle has been proposed \citep{Mohrhoff2014}.  
\begin{itemize}
\item[(I)] Whenever quantum mechanics requires us to add amplitudes, the distinctions we make between the alternatives correspond to nothing in the physical world.
\end{itemize}
This is a statement about the structure or constitution of the physical world, not a statement merely of our practical or conceptual limitations.

While the wave-function formulation stumps us with the dual problem of collapse and objectification, Feynman's formulation presents us with a question to which there is a clear answer. The reason why quantum mechanics requires us to add amplitudes is that the distinctions we make between the alternatives cannot be objectified (represented as real). The issue is not how measurement outcomes become objective but why distinctions we tend to make \emph{cannot} be considered objective.

\section{Identical particles}\label{sec:idpas}
Armed with a new interpretive principle, we set out to apply it to two paradigmatic setups, one concerning distinctions between \emph{regions of space}, the other concerning distinctions between \emph{things}. Applied to a two-slit experiment (or any two-way interferometer experiment, for that matter), (I)~tells us that the distinction we make between $p_L$ = ``the particle went through the left slit ($L$)'' and $p_R$ = ``the particle went through the right slit ($R$)'' corresponds to nothing in the physical world. If this distinction could be objectified, the particle would take either $L$ or $R$, in which case the interference fringes predicted whenever we are required to add amplitudes would not be seen. In \emph{some} sense the particle took both slits, but not in the sense of the conjunction $p_L\wedge p_R$, for one never sees that a particle emerges from the left slit \emph{and} from the right slit. To say that the particle went through both slits can only mean that it went through the union $L\cup R$ of the two slits without going through a particular slit and without being divided into parts that go through different slits.%
\footnote{Because we are here concerned with the particle's position, the question of parts does not arise, for a position isn't a thing that can have parts. It can only be indefinite or fuzzy.}
This confirms our earlier conclusion that physical space cannot be an intrinsically differentiated expanse, for if the parts of space defined by the slit were intrinsically distinct, a particle could not go through both slits in the sense just spelled out.%
\footnote{For a significantly more detailed presentation of this argument see \citep[Sec.~4]{Mohrhoff2014}.}

The consistency of interpretive principle (I) might be challenged. If the distinctions we make between alternatives do not correspond to objective differences in the physical world, how is it nevertheless possible for us to make these distinctions? The two-slit experiment should make this clear. There are two alternatives: the particle could go through $L$ or it could go through~$R$. The possibility of making this distinction does not depend on whether a measurement determining the slit taken by the particle is actually made. But only if this measurement is made can the distinction between $p_L$ and $p_R$ be objectified.

Applied to the propagation of a pair of identical particles (i.e., particles of the same type, lacking individualizing characteristics), (I)~tells us that the distinction we make between alternative identifications corresponds to nothing in the physical world. The elastic scattering of two identical particles may serve as an example. (A scattering process is called ``elastic'' if no particles are created or annihilated in the process.) Given two identical particles initially moving northward and southward, respectively, the probability of eventually finding one particle moving eastward and one particle moving westward takes the form
\be
\bigl|\braket{EW}{NS}\pm\braket{WE}{NS}\bigr|^2,\label{eq-EWNS}
\ee
where the sign depends on whether the particles are bosons or fermions. This expression can also be obtained by using the Born rule with the following initial and final states:
\be
\ket{\psi_i}=\frac1{\sqrt2}\bigl(\ket{NS}\pm\ket{SN}\bigr),\quad
\ket{\psi_f}=\frac1{\sqrt2}\bigl(\ket{EW}\pm\ket{WE}\bigr).
\ee
It is then readily seen why any evolving-states formulation of quantum mechanics requires the use of (anti)symmetrized particle states. If we were to use $\ket{AB}$ instead of the (anti)symmetrized product, we would introduce, in addition to the physically warranted distinction between ``the particle in $A$'' and ``the particle in $B$,'' the physically unwarranted distinction between the ``first'' or ``left'' particle and the ``second'' or ``right'' particle (in the expression $\ket{AB}$). This would be justified if the particles carried ``identity tags'' corresponding to ``left'' and ``right,'' in which case we would be required to add probabilities, not amplitudes. If the distinction between ``the particle in $A$'' and ``the particle in $B$'' is the only physically warranted distinction, the distinction between the  ``left'' particle and the ``right'' particle must be eliminated, and this is achieved by (anti)symmetrization.

Because in (\ref{eq-EWNS}) amplitudes are added rather than probabilities,  (I)~applies, and it tells us that the distinction we make between the alternative identifications

\medskip\centerline{$N=E,S=W$\quad or\quad $N=W,S=E$}

\medskip\noindent corresponds to nothing in the physical world. There is no answer to the question: ``Which outgoing particle is identical with which incoming particle?'' What does this tell us about the physical world? The answer I propose is simple, straightforward, but also highly counterintuitive. It is that the incoming particles (and therefore the outgoing ones as well) are \emph{one and the same entity}. The reason this answer is counterintuitive is that we tend to think of distinct parts of space as self-existent (rather than as being distinct only by virtue of being separately realized by different detectors), and that we tend to think of whatever is contained in distinct parts of space as separately and independently existing things. Apart from ``our habit of inappropriately reifying our successful abstractions'' \citep{Mermin2009}, this unexamined way of thinking about space and things appears to me to be one of the greatest obstacles to unraveling the magic and mystery of quantum mechanics.

Initially we observe two distinct properties, the property of moving northward and the property of moving southward, and subsequently we again observe two distinct properties, the property of moving eastward and the property of moving westward. We do not really observe two distinct things with two distinct properties---this is one ``two'' too many. What we observe is perfectly consistent with saying that, both initially and subsequently, we observe the same thing twice, with two distinct properties. Here as elsewhere, unanswerable questions tend to arise from false assumptions. The reason why ``Which outgoing particle is identical with which incoming particle?'' is unanswerable may well be that the question implicitly assumes, wrongly, that, initially as well as subsequently, there are two things.

But are there things over and above the initially and subsequently observed properties? In other words, are there things over and above measurement outcomes and their correlations? Are there particles? In order to throw light on these questions, let us consider a series of position measurements, and let us assume that each measurement yields exactly one outcome (i.e., each time exactly one detector clicks). What we then have, in addition to detector clicks and correlations between detector clicks, is evidence of a conservation law. If each time exactly two detectors click, we have evidence of the same conservation law, the conserved quantity being the number of simultaneous clicks (or the maximum number if the less than perfect efficiency of real-world detectors is taken into account). If this is the only conservation law in force (i.e., if the particles do not carry ``identity tags''), can we think of the system under study as being ``made of'' a fixed number of particles? The answer is No, for we cannot then think of these particles as re-identifiable individuals. In the quantum domain, individual identity is predicated on the existence of individualizing properties (``identity tags''), not on the distinguishability of things \emph{per se}---things considered in the absence of properties by which they can be distinguished. Considered in the absence of properties by which things can be distinguished, there is only one thing---the system as a whole.%
\footnote{There is a philosophical position according to which there are no things---there are only bundles of properties. Whenever we apply quantum mechanics to a physical system, there is, however, always one thing: the physical system as a whole. We may not be able to think of it as being ``made up'' of ``smaller'' things, but we are always in a position to think of it as the thing we study experimentally, and we are always able to attribute to it the properties we observe.} 

In the relativistic theory, the number of simultaneous detector clicks becomes another quantum observable, which can come out different every time it is measured. Regardless of whether this number is fixed or variable, however, the indistinguishability of identical particles implies a unity of the system as a whole that goes beyond the unity of ``a Many that allows itself to be thought of as a One'' (Cantor's definition of a set). Instead, we are in the presence of \emph{a One that allows itself to be thought of as a Many}---a ``thing \emph{per se}'' that takes on a fixed or variable number of positions each time its positions are measured. In our elastic scattering example, the two particles are a One that manifests itself twice. The incoming particles (and therefore the outgoing ones as well) are one and the same entity. What's more, there is no compelling reason to believe that this identity ceases when it ceases to have observable consequences owing to the presence of individuating properties. We are free to take the view that \emph{intrinsically} each particle is numerically identical with every other particle. What presents itself here and now with these properties and what presents itself there and then with those properties is one and the same entity.%
\footnote{I am not the first to put forth this preposterous idea. In his Nobel Lecture  on December 11, 1965, Feynman recalled: ``I received a telephone call one day at the graduate college at Princeton from Professor Wheeler, in which he said, `Feynman, I know why all electrons have the same charge and the same mass.' `Why?' `Because, they are all the same electron!'\,''}
In what follows I shall call it ``Being.'' If you prefer any other name, be my guest.%
\footnote{There is an extensive literature on the subject of individuality in quantum theory. See \citet{French} for an overview and \citet{FrenchKrause} for a comprehensive review. French sums up the situation by stating that quantum mechanics is ``compatible with two distinct metaphysical `packages,' one in which the particles are regarded as individuals and one in which they are not.'' I rather agree with \citet{Esfeld}, who does not consider it ``a serious option to regard quantum objects as possessing a primitive thisness (haecceity) so that permuting these objects amounts to a real difference.''}

\section{Manifestation}\label{Manifestation}
Another reason it is so hard to make sense of the quantum theory is that it answers a question we are not in the habit of asking. Instead of asking what the ultimate constituents of matter are and how they interact and combine, we need to broaden our repertoire of explanatory concepts and ask: \emph{how are forms manifested?} This question, too, has a simple, straightforward, and counterintuitive  answer \citep[Sec.~9]{Mohrhoff2014}: The shapes of things resolve themselves into reflexive spatial relations. By this I mean self-relations---relations between Being and the self-same Being, relations between numerically identical relata. By entering into (or by entertaining) reflexive spatial relations, Being gives rise to (or supports)
\ben
\item what looks like a multiplicity of relata if the reflexive quality of the relations is ignored, and 
\item what looks like a substantial expanse if the spatial quality of the relations is reified.
\een
Because the relations are reflexive, the multiplicity of the relata is apparent rather than real.%
\footnote{Does this mean that the material world is unreal, as some illusionistic philosophies assert? By no means, for the material world owes its existence to an intrinsically undifferentiated Being and a multitude of reflexive relations, and these are real.}
And because physical space, insofar as it consists of anything, consists of spatial relations, the spatial quality belongs to the relations themselves; they do not owe it to a substantial expanse. But if physical space is the set of all reflexive relations, the shapes of things are subsets of this set; they are particular sets of spatial relations.

The view put forward here goes farther in relationism---the doctrine that space and time are a family of spatial and temporal relations holding among the material constituents of the universe---in that it also affirms that the ultimate material constituents are formless. While fundamental particles are routinely described as pointlike, what is meant is that they lack internal structure. Lack of internal structure is suggested by the scale-invariance of a particle's effective cross-section(s) in scattering experiments with probe particles that are themselves pointlike in this sense, but can be verified only down to the de Broglie wavelength of the probe particles. There can therefore be no evidence of absence of internal structure, let alone evidence of a literally pointlike form. For further reasons why fundamental particles ought to be conceived as formless, see \citep[Sect.~9]{Mohrhoff2014}. So conceived, the shapes of things resolve themselves into sets of spatial relations between formless and numerically identical relata. The truism that the universe lacks a position because it lacks \emph{external} spatial relations thus has a fitting complement: a fundamental particle lacks a form because it lacks \emph{internal} spatial relations.

Since it is necessary to distinguish between (i)~properties that only exist if and when they are measured and (ii)~measurement-independent properties that are capable of indicating outcomes, the distinction between a classical or macroscopic domain and a non-classical or quantum domain is amply justified. But how can we understand the relation between the two domains, beyond the linguistic necessity of speaking about the quantum domain in terms of correlations between macroscopic events, which was stressed by \citet{Bohr1934}? The answer I propose is that the distinction between the two domains is essentially a distinction between the \emph{manifested world} and its \emph{manifestation}.

Atoms and subatomic particles must in some way be responsible for the existence of the objects that populate the familiar world of everyday experience. But since the kinematical properties of microphysical objects---their positions, momenta, energies, etc.---only exist if and when they are indicated by the behavior of macroscopic objects, they cannot play the role of interacting constituent parts. Macroscopic objects cannot be said to be \emph{made of} microphysical ones.%
\footnote{When physicists reflect on the motives for their research, they nonetheless often claim (especially on TV, in press releases, and in grant applications) that their aim is to discover the elementary building blocks of the universe and the processes by which these interact with each other---a rather schizoid state of affairs.}
A broader concept is needed to understand the relation between the two domains, such as the concept of \emph{manifestation}. How, then, is this concept to be understood? To begin with, since the manifestation of the world includes the manifestation of both space and time, it cannot be conceived as a process that takes place in space and time. We keep looking for the origin of the universe at the beginning of time, but this is an error of perspective. I propose to identify the origin of the universe with the Being that was introduced in the previous section, a Being intrinsically undifferentiated, transcendent of spatial and temporal distinctions. And I propose to look upon the manifestation of the world as a transition from the undifferentiated state of Being to a state that allows itself to be described in the classical language of interacting objects and causally related events---a transition from absolute unity to the multiplicity of the macroworld. If there is a metaphysical message that quantum mechanics is trying to convey to us, it concerns, I maintain, this transition. The theory thereby reverses the explanatory arrow of common sense and folk physics (a.k.a. classical physics): instead of attempting to explain wholes in terms of interacting parts, it suggests how the multiplicity of the world emerges from an intrinsically undifferentiated Being. 

The transition from the absolute unity of Being to the multiplicity of the macroworld passes through several stages. Across these stages, the world's differentiation into distinguishable regions of space and distinguishable objects with definite properties is gradually realized. There is a stage at which Being presents itself as a multitude of formless particles. This stage is probed by high-energy physics and known to us through correlations between the counterfactual clicks of imagined detectors, i.e., in terms of transition probabilities between in-states and out-states. There are stages that mark the emergence of form, albeit a type of form that cannot yet be visualized. The forms of nucleons, nuclei, and atoms can only be mathematically described, as probability distributions over abstract spaces of increasingly higher dimensions. At energies low enough for atoms to be stable, it becomes possible to conceive of objects with fixed numbers of components, and these we describe in terms of correlations between the possible outcomes of unperformed measurements. The next stage---closest to the manifested world---contains the first objects with forms that can be visualized---the atomic configurations of molecules. But it is only the final stage---the manifested macroscopic world---that contains the actual detector clicks and the actual measurement outcomes that have made it possible to discover and study the correlations that govern the quantum domain.

A further question presses itself upon us. Since most measurable quantities only exist, or only have values, if and when they are measured, we cannot account for the properties of macroscopic objects in terms of the properties of microphysical objects. How, then, can atoms and subatomic particles be instrumental in the manifestation of the macroworld? How can this manifestation be understood without reference to the possessed properties of microphysical objects? The answer is that atoms are known to us through correlations between measurement outcomes,%
\footnote{Generations of students have been puzzled by the special role that the $z$~axis plays in descriptions of the stationary states of atomic hydrogen. If a stationary state were to describe an atom as it is by itself, the question would have no answer. A stationary state, however, is a probability algorithm predicated on a particular preparation. In describing the atom's stationary states we \emph{assume} that a particular component of its angular momentum has been measured, along with its energy and its total angular momentum.}
while subatomic particles are known to us through correlations between detector clicks. Their respective roles in the manifestation of macroscopic objects can be described in terms of conditional propositions expressing correlations, and thus without referring to actually measured (and thus actually possessed) properties.

Many of the mysteries surrounding quantum mechanics become clear in this light. Why, after all, is the general theoretical framework of contemporary physics a probability calculus, and why are the probabilities assigned to measurement outcomes? If quantum mechanics concerns a transition through which the differentiation of the world into distinguishable objects and distinguishable regions of space is gradually realized, the question arises as to how the intermediate stages are to be described---the stages at which the differentiation is incomplete and the distinguishability between objects or regions of space is only partially realized. The answer to this question is that whatever is not completely distinguishable can only be described by assigning probabilities to what is completely distinguishable, namely to the different possible outcomes of a measurement. What is instrumental in the manifestation of the world can only be described in terms of what happens in the manifested world, or else in terms of correlations between events that happen or could happen in the manifested world.

It is also worth stressing that the indeterminism of quantum mechanics does not fundamentally consist in the existence of unpredictable changes disrupting a predictable evolution, as the myth of evolving quantum states suggests. Instead, it is rooted in the respective indeterminacies of the various stages of manifestation. The unpredictability of measurement outcomes is merely the consequence of underlying indeterminacies, which evince themselves through unpredictable transitions in the values of outcome-indicating positions (the proverbial ``measurement pointers'').

Quantum mechanics thus presents us with a so far unrecognized kind of cau\-sal\-ity---unrecognized, I believe, within the scientific literature albeit well-known to metaphysics, inasmuch as the general philosophical pattern of a single world-essence (``Being'') manifesting itself as a multiplicity of individual things is found throughout the world.%
\footnote{Some of its representatives in the Western hemisphere are the Neoplatonists, John Scottus Eriugena, and the German idealists. The quintessential Eastern example is the original (pre-illusionist) Vedanta of the Upanishads \citep{Phillips,SA1,SA3}.}
For the causality of the process of manifestation must be distinguished from its more familiar temporal cousin, which links states or events across time or spacetime. The usefulness of the latter is confined to the world drama; it plays no part in setting the stage for it. It helps us make sense of the manifested world as well as of the cognate world of classical physics, but it throws no light on the process of manifestation nor on the quantum correlations that are instrumental in the process.

While an atemporal causality does not involve a temporal sequence, process, or transition, there are stages by which the differentiation of the world into distinguishable objects and distinguishable regions is gradually realized. The coexistent stages of this gradual realization can be placed along a dimension of logical space that is neither temporal nor spatial, and they can be viewed in a logical sequence, as a transition from undifferentiated unity to the multiplicity of the manifested world, via numerically identical particles, non-visualizable atoms, and partly visualizable molecules. We do, in fact, take much the same liberty when we conceive, as we routinely do, of temporal succession as if it were another spatial dimension. If we have the right to visualize time as a dimension of a 4-dimensional expanse,  then we also have the right to imagine an atemporal causal arrow and to attribute the existence of the world to an atemporal process of manifestation.

It may be of interest to compare the manifestation of the macroworld with the traditional philosophical concept of the emergence of the Many from a One. In classical metaphysics, this emergence was conceived as running parallel to predication: an immaterial essence or predicable universal becomes instantiated as an impredicable material individual. This instantiation, moreover, used to be conceived in the framework of Platonic--Aristotelian dualism, which postulates an instantiating medium (matter and/or space) in or by which the essences or universals get instantiated. The manifestation of the macroworld, by contrast, requires no separate medium and implies no dualism. All that is required is the realization of spatial relations. Being may be said to manifest the macroworld \emph{within} itself---after all, the macroworld is manifested with the help of \emph{reflexive} relations---rather than in something other than itself.

\section{Quantum nonlocality}\label{nonlocality}
\bq
\emph{For those interested in the fundamental structure of the physical world, the experimental verification of violations of Bell's inequality constitutes the most significant event of the past half-century. In some way our basic picture of space, time, and physical reality must change. These results, and the mysteries they engender, should be the common property of all who contemplate with wonder the universe we inhabit.} \citep[p.~4]{Maudlin}
\eq
In his seminal paper of 1964, Bell showed that the principle of local causes (also called Einstein locality) was incompatible with quantum mechanics---a result that was hailed by \citet{Stapp} as ``the most profound discovery of science.'' In his famous cat paper, Schr\"odinger noted that ``Measurements on separated systems cannot directly influence each other---that would be magic.'' Bell showed that the magic was real. His conclusion was that
\bq
In a theory in which parameters are added to quantum mechanics to determine the results of individual measurements, without changing the statistical predictions, there must be a mechanism whereby the setting of one measuring device can influence the reading of another instrument, however remote. \citep{Bell1964}
\eq
The reason Bell examined deterministic theories, in which unobservable parameters are added to quantum mechanics, was not that he was averse to indeterminism but that deterministic theories were the only hope for retaining locality, a hope that was dashed by him for good. While in an indeterministic theory locality is doomed by perfect correlations, in a deterministic theory it is doomed by Bell's inequality. No matter whether measurement outcomes are determined by hidden parameters or stochastic, there must be a causal connection between the setting of one apparatus and the reading of another, and if an inertial frame exists in which the two events---the setting of the first apparatus and the reading of the second---are simultaneous, the connection will be superluminal. It is then called ``nonlocal'' because it can be accounted for neither by a direct influence propagating locally from one event to the other nor by a common cause in the intersection of the past light cones of the two events.

The causality associated with the atemporal process of manifestation presents the nonlocality of quantum mechanics in a new light. I contend that quantum mechanics violates locality for the same reason that the manifestation of the world cannot be explained by processes that connect events across spacetime, spacetime being an aspect of the finished product, the manifested world. The  atemporal process by which Being enters into reflexive relations and matter and space come into being as a result, is the nonlocal event \emph{par excellence}. Depending on one's point of view, it is either coextensive with spacetime (i.e., completely delocalized) or ``outside'' of spacetime (i.e., not localized at all). Occurring in an anterior relation to space and time, it is the common cause of all correlations, not only of the seemingly inexplicable ones between simultaneous events in different locations but also of the seemingly explicable ones between successive events in the same location.

I also contend that the diachronic quantum correlations between events in timelike relation are in reality as mysterious and inexplicable (in terms of across-space-and-time causality) as the synchronic quantum correlations between events in spacelike relation. The notion that the former can be explained causally, through a dynamical evolution that connects outcomes of measurements made on the same system at different times, is an illusion. This kind of evolution misrepresents the dependence of probabilities on the times of the measurements to the outcomes of which they are assigned. We know as little of a physical process by which an event here and now contributes to determine the probability of a \emph{later} event \emph{here} as we know of a physical process by which an event here and now contributes to determine the probability of a \emph{distant} event \emph{now}.

The nonlocality implied by Bell's theorem and similar no-go theorems thus is but the most salient symptom of a much deeper and more general nonlocality. It is the nonlocality of that intrinsically undifferentiated Being, one with every fundamental particle, which manifests the world by entering into (or entertaining) reflexive spatial relations. It is the nonlocality of the process of manifestation, which yields an apparent locality (i.e., amenability to local explanation) only in its final outcome, the manifested world.

Quantum physics does not explain ``how nature does it.'' As \citet[p.~78]{FeynmanQED} put it, ``[t]here are no `wheels and gears' beneath this analysis of Nature.'' The theory only explains---via conservation laws---why certain things \emph{won't} happen. This is exactly what one would expect if the force that has set the stage for the drama of an evolutionary unfolding of life%
\footnote{A hundred years ago it seemed obvious to many that life could not have emerged from utterly lifeless matter, just as today it seems obvious to many that experience cannot have emerged from utterly non-experiential matter. Yet today no one appears to seriously doubt that life did emerge from utterly lifeless matter; the seemingly insuperable ``hard problem of life'' simply dissolved. So why should it not be the same with the ``hard problem of consciousness'' \citep{Chalmers}, a hundred years from now? As \citet{Strawson} has pointed out, one cannot draw such a parallel unless one considers life completely apart from conscious experience. If consciousness is essential to life, as it may well be, one cannot reduce life to physics via chemistry if one cannot reduce consciousness along with it.}
and mind were an infinite (unlimited) force operating under self-imposed constraints. In that case we would have no reason to be surprised or dismayed by the impossibility of explaining the quantum-mechanical correlation laws in terms of physical mechanisms. For it would be self-contradictory to invoke a physical mechanism to explain the working of an infinite force. What would need explaining is why this force works under the particular constraints that it does. Where the efficient causality of the atemporal process of manifestation fails, final causality takes over. As was suggested in the Introduction, what is possible is to explain why the laws of physics have the specific form that they do \citep{Mohrhoff2002,Mohrhoff2009b,Mohrhoff2011}.

\section{Postscript: consciousness and Being }\label{sca}
If quantum states are regarded as evolving physical states, and if their unitary ``dynamics'' is taken to be their natural mode of evolution, then collapse and objectification appear to be preternatural interventions into the course of Nature, as was pointed out in Section~\ref{sec:semantic}. Because the obvious possible perpetrator of such interventions is the consciousness of the observer, \citet{vonNeumann} came to argue that ``we must always divide the world into two parts, the one being the observed system, the other the observer.'' But neither are quantum states evolving physical states, nor do measurement outcomes exist merely in the minds of observers. A measurement outcome is as observer-independent as the change in the value of any macroscopic position from which it can be inferred (regardless of whether anyone is around to make the inference). Contrary to claims made by \citet{Stapp2011} and others, quantum mechanics \emph{can} be understood without reference to conscious observers---to the extent that it can be understood.

This does not mean that quantum mechanics cannot throw light on the mysterious presence of consciousness in what appears to be a material universe. To my mind, what holds the key to this conundrum is the self-identical Being that constitutes every particle in existence. The root of consciousness is not to be found in the manifested world, nor in the process of manifestation, but in that which manifests the world. For Being does not simply manifest the world; it manifests the world \emph{to itself}. Being relates to the world not only as the substance that constitutes it but also as the consciousness that contains it. It is at once the single substance by which the world exists, and the ultimate self or subject for which it exists.

 How, then, are our conscious selves related to this ultimate self or subject? This question has been answered in considerable detail and on a solid experiential foundation by the Indian philosopher (and freedom fighter, and mystic) Sri Aurobindo \citep{Heehs}. In keeping with a more than millennium-long philosophical tradition \citep{Phillips}, \citet{SA5} posits an Ultimate Reality whose intrinsic nature is (objectively speaking) infinite Quality and (subjectively speaking) infinite Delight. This has the power to manifest its inherent Quality/Delight in finite forms, and the closest description of this manifestation is that of an all-powerful consciousness creating its own content.

In the native poise of this consciousness, its single self is coextensive with its content and identical with the substance that constitutes the content. A first self-modification of this \emph{supramental} consciousness leads to a poise in which the one self adopts a multitude of standpoints, localizing itself multiply within the content of its consciousness and viewing the same in perspective. It is in this secondary poise that the dichotomy between subject and object, or self and substance, becomes a reality. 

Probably the most adequate description of the process by which the one self assumes a multitude of standpoints is that of a multiple concentration of consciousness. A further self-modification of the original creative consciousness occurs when this multiple concentration becomes exclusive. We all know the phenomenon of exclusive concentration, when consciousness is focused on a single object or task, while other goings-on are registered subconsciously, if at all. A similar phenomenon transforms individuals who are conscious of their essential mutual identity into individuals who have lost sight of this identity and, as a consequence, have lost access to the supramental view of things. Their consciousness is mental, which means not only that it belongs to a separate individual but also that it perceives or presents the world as a multitude of separate objects.

\citet{Kafatos-spooky}  recently expressed the view that ``The main issue is not why our universe is nonlocal and entangled but rather how it appears as being made of distinct, physical objects.'' According to him, the fundamental issue is how our consciousness ``takes on limited characteristics, divides the infinite and in the end creates a limited, divided, physicalist view'' of reality. That, indeed, is the question, and the answer is: by a multiple exclusive concentration of the creative consciousness that is one with Being. The belief that ``things claim an existence independent of one another'' whenever they ``lie in different parts of space'' \citep{Einstein48} gives expression to the mind's way of dealing with reality, not to the supramental truth of things. I agree with Kafatos that consciousness ``is primary, rather than derived through blind processes as current materialist approaches claim.'' Consciousness is primary because Being is primary, and because at its origin consciousness is one with Being. But I do not see the quantum-mechanical phenomena of coherence, entanglement, and nonlocality as (more or less direct) evidence that ``the universe \dots\ is basically living and conscious,'' as Kafatos does. As I see it \citep{Mohrhoff2002,Mohrhoff2009b, Mohrhoff2011}, the well-tested laws of physics (including the aforesaid phenomena) are largely consequences of the existence of  ``ordinary'' objects as defined in note \ref{oo}. As long as the existence of such objects is taken for granted, there is no reason to implicate consciousness. It is only when we address the purely metaphysical question why such objects are composed of finite numbers of objects lacking spatial extent that  consciousness enters the picture.  

Here is how. If one of the characteristics of mind is the ability to devise plans for action and to generate ideas for creation, one of the characteristics of life is the ability to carry out such plans and to execute such ideas. Mentally conscious living beings come into existence not \emph{only} by an evolution from seemingly lifeless and unconscious matter but also, and in the first place, by a multiple exclusive concentration of the creative consciousness that is one with Being. What happens if this multiple exclusive concentration is carried to its logical conclusion? What happens then is the coming into  being of a world whose inhabitants lack not only the ability to generate ideas but also the power to execute them. And since this power is responsible for the existence of individual forms, the result is a world of formless individuals, which we are used to calling ``fundamental particles.'' This is how the original creative consciousness came to be ``involved'' in mind, how mind came to be ``involved'' in life, and how life came to be ``involved'' in formless particles. And that is why formless particles are capable of evolving life, why life is capable of evolving mind, and why mind may eventually evolve into the supramental consciousness, which is Being's power of self-determination.

The action of this supramental consciousness is primarily qualitative and infinite and only secondarily quantitative and finite. Essentially, mind is the agent of the supermind's secondary, quantifying, and delimiting action. But when it is separated in self-awareness from its supramental parent consciousness, as it is in us, it not only divides \emph{ad infinitum} but also takes the resulting multiplicity for the original truth or fact. This is why we tend to construct reality from the bottom up, on an intrinsically and completely differentiated space or space-time, out of locally instantiated physical properties, or else by aggregation, out of a multitude of individual substances. And that, at bottom, is why making sense of quantum mechanics is so hard. For, as we found in Section~\ref{sec:semantic}, the spatial differentiation of the physical world does not go ``all the way down.'' Reality is structured by a progressive self-differentiation of Being that does not ``bottom out,'' and this makes it impossible to construct reality from the bottom up.

\bigskip\noindent\textbf{Ulrich J. Mohrhoff} studied physics at the University of G\"ottingen and at the Indian Institute of Science in Bangalore. He now teaches at the Sri Aurobindo International Centre of Education (Pondicherry, India). He is the author of the textbook \emph{The World According to Quantum Mechanics: Why the Laws of Physics Make Perfect Sense After All} (Singapore: World Scientific, 2011). His research interests are the Foundations of Quantum Physics, the Philosophy of Mind, and the Interface between Contemporary Physics and Indian Philosophy.

\end{document}